\documentclass[aps,prc,twocolumn,preprintnumbers,amsmath,showpacs,floatfix,nofootinbib]{revtex4-1}
\usepackage{bm,graphicx}

\newcommand{\dd}{\mathrm{d}}

\begin{document}

\preprint{BI-TP 2014/024}

\title{Kinetic freeze out from an anisotropic fluid in high-energy heavy-ion collisions: particle spectra, Hanbury Brown--Twiss radii, and anisotropic flow}

\author{Nicolas Borghini} \email{borghini@physik.uni-bielefeld.de}
\author{Steffen Feld} \email{s.feld@physik.uni-bielefeld.de}
\author{Christian Lang} \email{chlang@physik.uni-bielefeld.de}
\affiliation{Fakult\"at f\"ur Physik, Universit\"at Bielefeld, Postfach 100131, D-33501 Bielefeld, Germany}

\date{\today}

\begin{abstract}
  Dissipative relativistic fluid-dynamical descriptions of the extended fireball formed in high-energy heavy-ion collisions are quite successful, yet require a prescription for converting the fluid into particles.   
  We present arguments in favour of using a locally anisotropic momentum distribution for the particles emitted from the fluid, so as to smooth out discontinuities introduced by the usual conversion prescriptions. 
  Building on this ansatz, we investigate the effect of the asymmetry on several observables of heavy ion physics. 
\end{abstract}

\pacs{25.75.Ld, 12.38.Mh}

\maketitle

\section{Introduction}

A large amount of the dynamical properties of the fireball created in high energy collisions of heavy nuclei---be it at the Brookhaven Relativistic Heavy Ion Collider (RHIC) or at the CERN Large Hadron Collider (LHC)---can be described to a good approximation within the framework of relativistic fluid dynamics (see Ref.~\cite{Huovinen:2013wma} for a critical review). 
The relevant equations of motion have to be supplemented with appropriate initial conditions for the continuous medium and with a recipe for the end of its evolution, namely the conversion of the fluid into particles~\cite{Huovinen:2012is}. 
Our focus in this work will be on the latter point and on how adopting a new ansatz for the transition can help mitigate a few issues in the usual approaches.

After their emission from the fluid---in which the mean free path is assumed to be very small---, the particles can be left to propagate freely, i.e.\ they at once acquire an infinitely large mean free path: 
the conversion step is the so-called (kinetic) freeze out, after which the particle momenta no longer evolve. 
Alternatively, the emitted particles can be fed into an ``afterburner'' that implements a set of transport equations for the various particle species and thereby ensures a more gradual change of the mean free path (see Ref.~\cite{Hirano:2012kj} for a recent overview).
The fluid--particle transition is then a switch between model descriptions, rather than a physical phenomenon. 

Irrespective of the subsequent fate of the particles, their emission from the fluid is often modelled in a similar way.  
For that reason, we shall generally for the sake of simplicity refer to the conversion process as ``freeze out'', although the actual decoupling occurs later when the particles are evolved with a transport code. 

Most existing studies follow some version of the Cooper--Frye prescription~\cite{Cooper:1974mv}: 
the fluid breaks up, more or less suddenly, when reaching a three-dimensional hypersurface $\Sigma$---sometimes replaced by a thin four-dimensional shell, to mitigate the inherent abruptness of the description---defined by some a priori criterion, like a constant temperature or energy density. 
At each point on the freeze-out hypersurface, particles are emitted with a given phase space distribution $f(\mathsf{x},\mathsf{p})$.%
\footnote{We denote four-vectors in sans serif font and three-vectors in boldface;  for the metric we adopt the mostly-minus convention.} 
Integrating over the whole hypersurface, the resulting invariant spectrum of the emitted particles of type $i$ reads
\begin{equation}
\label{Cooper-Frye}
E_{\bm{p}}\frac{\dd^3N_i}{\dd^3\bm{p}} = 
\frac{g_i}{(2\pi)^3}\!\int_\Sigma\!f_i(\mathsf{x},\mathsf{p})\, p^\mu\dd^3\sigma_\mu(\mathsf{x}).
\end{equation}
In this expression, we have taken into account the fact that the phase space distribution depends on the particle species, especially its bosonic or fermionic nature, and we explicitly factorized out the particle degeneracy factor $g_i$. 

The phase space occupation factor in the Cooper--Frye formula is chosen so as to ensure the conservation of energy, momentum, and charges---if any---across the freeze-out hypersurface. 
Accordingly, $f_i(\mathsf{x},\mathsf{p})$ is usually taken to be the equilibrium thermal distribution---which is appropriate for a perfect fluid---or a near-equilibrium distribution including ``correction terms'' that match the stress energy tensor of a dissipative fluid. 
Determining these corrections either from pure  theory~\cite{Teaney:2003kp,Dusling:2007gi,Denicol:2009am,Monnai:2009ad,Dusling:2009df,Pratt:2010jt,Dusling:2011fd,Teaney:2013gca,Molnar:2014fva} or within more phenomenological data-driven approaches~\cite{Luzum:2010ad,Lang:2013oba} is an ongoing effort.
In any case, it is always implicitly assumed that the decoupling medium is not far from local equilibrium, so that dissipative effects remain small. 

Thus, the occupation factors $f_i$ at freeze out considered in the literature are, up to small corrections, isotropic in the fluid local rest frame, reflecting the assumption of (near) local thermal equilibrium. 
Accordingly, $f_i$ depends on position only through the corresponding dependence of thermodynamic variables, namely the flow velocity $\mathsf{u}(\mathsf{x})$ and its gradients, the freeze-out temperature $T_{\mathrm{f.o.}}(\mathsf{x})$, and possibly the chemical potential $\mu_i(\mathsf{x})$. 

In this work, we shall depart from this local isotropy of $f_i$ and assume instead a locally asymmetric momentum distribution at decoupling. 

Before proceeding any further, let us mention that the existence of some local momentum anisotropy at freeze out was already considered in Ref.~\cite{Rybczynski:2012ee}.
As will become clear in the following section, the anisotropy we are interested in is of a different kind, reflecting the dissimilar underlying motivation. 
Nevertheless, some of the findings of Ref.~\cite{Rybczynski:2012ee} naturally translate into similar results in our case.

\section{Motivation}

The sudden-decoupling scenario embodied in the Cooper--Frye formula~\eqref{Cooper-Frye} aims at gluing together two rather different descriptions. 
The mismatch of the models is obvious if the fluid freezes out into free-streaming particles, as exemplified by the jump of the Knudsen number from very small to very large values. 
Even when the Cooper--Frye prescription is used to switch from a dissipative fluid to a collection of interacting hadrons, there remain issues~\cite{Huovinen:2012is,Hirano:2012kj}. 
An often mentioned problem is the existence of sectors of the conversion hypersurface $\Sigma$ where $\dd\sigma_\mu(\textsf{x})\,\dd\sigma^\mu(\textsf{x})<0$, which can locally lead to negative contributions to the Cooper--Frye integral. 
Cures to this issue have been proposed (see e.g. Ref.~\cite{Grassi:2004dz} and references therein), which themselves remain incomplete since they introduce discontinuities across $\Sigma$ either in the stress energy tensor or in the velocity. 
These shocks are however artefacts of the modelling, not physical ones. 

Another issue of the usual sudden freeze out recipe is the sensitivity of the observables computed with the emitted particles, in particular their spectra, to the parameters in the Cooper--Frye formula. 
This is in our eyes a rather crucial point: it means that the matching between a ``microscopic'' approach and a long-wavelength effective theory thereof, namely the kinetic modelling in terms of particles and the fluid-dynamical description, depends significantly on the parameter that separates them, which makes the whole procedure questionable.
 
A strong theoretical incentive for developing and investigating new approaches to the modelling of decoupling at the end of the dynamics of heavy-ion collisions is thus to obtain a description which interpolates between the hydrodynamic and ballistic regimes in a smoother manner than the usual prescriptions.

A possible way out of the problem is to drop the assumption of a sudden freeze out in favour of a continuous one~\cite{Grassi:1994nf&ng,Sinyukov:2002if}. 
However, in the current implementations of this approach, the particles decoupling from the fluid do not reinteract with each other afterwards. 
This again implies for each particle a sudden transition from a vanishingly small to an infinitely large mean free path---where the latter is viewed somewhat abusively as the average length that a given particle is likely to travel in its next step---, which is again unsatisfactory, even though this does not happen at once for the whole fluid. 

Despite its deficiencies, the ``naive'' Cooper--Frye formula remains attractive because of its simplicity, which makes it easier to test novel ideas.
In order to ensure a better transition between the fluid and particle description, it seems desirable to ``twist'' one of the models or both, so as to bring them closer to each other. 
In this spirit, we suggest that anisotropic hydrodynamics~\cite{Florkowski:2010cf,Martinez:2010sc,Ryblewski:2013jsa} can improve the smoothness of the transition between the continuous and particle frameworks. 
As we shall demonstrate in next section, this ansatz helps alleviating the sensitivity to the freeze-out temperature $T_{\mathrm{f.o.}}$: 
Introducing new control parameters, namely those governing the anisotropy of the phase-space distribution at decoupling, widens the possible range of values for $T_{\mathrm{f.o.}}$.
In the remainder of this section, we list a few arguments in favour of distorting the particle distribution at freeze out. 

First, in the context of heavy-ion physics there is an obvious analogy with the advocated use of anisotropic hydrodynamics at early stages of the medium evolution, to ease the transition from the locally asymmetric energy-momentum tensor of the fields left by the colliding nuclei to the almost isotropic tensor needed to apply usual hydrodynamics consistently. 
In the early evolution stage, the phase-space distribution is deformed along the axis of the nucleus--nucleus collision ($z$-axis), while in the case we are interested here we do not expect such a global direction for the anisotropy. 

As a matter of fact, our second incentive to resort to a possibly strongly anisotropic freeze-out distribution is the observation of a similar asymmetry, parametrized as two different translation temperatures along the streamlines and perpendicular to them, in hypersonic nonrelativistic flows~\cite{Hamel:1966}. 
These findings help us specify the kind of anisotropy we want to consider hereafter. 
Let us for simplicity focus on particle emission around midrapidity, so as to discard any anisotropy along the $z$-direction on symmetry grounds. 
Far from the fluid, each particle will tend to fly away radially, as implied by the simultaneous conservation of angular momentum and (kinetic) energy.
The dispersion of the momentum components transverse to the radial direction will thus be much smaller than that of the radial component.

Eventually, a third argument for assuming a deformed particle distribution is that such an anisotropy was actually found for post-freeze-out distributions arising from the decoupling through time-like portions of freeze-out hypersurfaces~\cite{Gorenstein:1983bu,Molnar:2005gy}.

Accordingly, we conclude that it would be helpful to adopt in the Cooper--Frye picture a freeze-out distribution which is already deformed, with a larger mean squared momentum along the radial direction. 
That is, adopting the Cartesian (out, side, long) system of femtoscopic studies, we assume a larger pressure along the local ``out''-direction than in the sidewards and longitudinal directions. 

In the present paper, this asymmetry is admittedly a mere assumption, motivated by the observations in non-relativistic studies in which freeze out happens when the local particle distribution has a sizeable anisotropy in momentum space, and by the incentive to have a smoother transition between the fluid and particle descriptions. 
The actual functional form of the phase-space distribution at freeze out, as well as the size of the parameters measuring the anisotropy, should emerge from a detailed kinetic description of the decoupling process~\cite{BFL_inprep}. 
In next section, we shall postulate such a form and examine the change induced by the momentum-space asymmetry on various observables of heavy-ion collisions. 

Note that the anisotropy we consider hereafter differs from that considered in Ref.~\cite{Rybczynski:2012ee}, in which the distribution is assumed to be distorted along the $z$-axis, as a remnant of the distortion along that direction in the initial state of the nucleus--nucleus collision. 
Both deformations can naturally be present at once, yet our purpose here is to examine the influence of a larger radial-momentum dispersion, so that we keep the pressures in the side- and long-directions equal.

\section{Effect of the local anisotropy on observables}
\label{s:effect_xi}

Let us assume for the phase-space distribution at decoupling of a particle species with mass $m$ a Romatschke--Strickland-like profile~\cite{Romatschke:2003ms}, namely
\begin{equation}
\label{f_an_LR}
f_{\!\!\!\!\begin{array}{c} \textrm{\scriptsize an.} \\[-2mm] \textrm{\tiny (lrf)} \end{array}\!\!\!\!}(\textsf{x},\textsf{p};\Lambda,\xi) = 
  \bigg[\exp\!\bigg(\!\frac{\sqrt{m^2+\bm{p}^{\prime 2} + 
  	\xi(\textsf{x})_{} p_{\textrm{out}}^{\prime 2}}}{\Lambda(\textsf{x})}\bigg) \mp 1 \bigg]^{-1}, 
\end{equation}
where $p'_{\textrm{out}}$ denotes the out component of the particle momentum $\bm{p}^{\prime}$ with respect to the local rest frame (lrf) of the fluid at position $\textsf{x}$. 
$\Lambda$, which generalizes the temperature, characterizes the scale over which the particle momentum takes significant values. 
As hinted at by the notations, both $\Lambda$ and the anisotropy parameter $\xi$ depend a priori on position and the particle type. 
Hereafter they will be treated as parameters, and for simplicity taken as constant over the freeze-out hypersurface $\Sigma$. 

The anisotropy parameter $\xi$ must be larger than $-1$, to ensure the positivity of the expression under the square root.
In order to obtain a larger pressure along the radial direction than perpendicular to it, $\xi$ should be negative. 

To test the influence of the momentum anisotropy in Eq.~\eqref{f_an_LR}, we assume some specific freeze-out flow profile and hypersurface $\Sigma$. 
We thus let the fluid decouple at a constant proper time $\tau_{\textrm{f.o.}}$ on an longitudinally infinite, azimuthally symmetric cylinder of radius $R$. 
Taking as coordinates in the laboratory frame the proper time $\tau$, space-time rapidity $\varsigma$, and cylindrical coordinates $r,\phi$, we assume for the fluid velocity on $\Sigma$ a generalised blast wave-like profile, namely~\cite{Siemens:1978pb,Huovinen:2001cy}
\begin{equation}
\label{blastwave}
u^r(r,\phi) = \bar{u}_{\max}\frac{r}{R} \bigg( 1 + 2\sum_n V_n\cos n\phi\bigg)
\end{equation}
for the radial coordinate, $u^\phi = u^\varsigma = 0$ in the azimuthal and $\varsigma$ directions, and eventually $u^\tau=\sqrt{1+(u^r)^2}$. 
With this choice, the phase-space occupation factor~\eqref{f_an_LR} reads, when expressed in the laboratory frame
\begin{widetext}
\begin{equation}
\label{f_an_lab}
f_{\textrm{an.}}(\textsf{x},\textsf{p};\Lambda,\xi) = 
  \bigg[\exp\!\bigg(\!\frac{\sqrt{[p^\tau u^\tau(\textsf{x}) - p^r u^r(\textsf{x})]^2 + 
    \xi[p^r u^\tau(\textsf{x}) - p^\tau u^r(\textsf{x})]^2}}{\Lambda}\bigg) \mp 1 \bigg]^{-1}. 
\end{equation}
\end{widetext}

Under these assumptions, we can numerically compute the Cooper--Frye integral, from which we can obtain the transverse momentum spectrum, Hanbury Brown--Twiss (HBT) radii $(R_{\rm out}, R_{\rm side}, R_{\rm long})$~\cite{Bertsch:1988db,Pratt:1990zq}, and the anisotropic flow coefficients $v_n$.
We shall focus on pions ($m=140\;$MeV) produced at midrapidity.  

We first present results obtained with fixed values of the ``effective temperature'' $\Lambda= 150\;$MeV and of the parameters of the blast wave profile: $\tau_{\textrm{f.o.\!}} = 7.5\;$fm/$c$, $R=10\;$fm, $\bar{u}_{\max} = 1$, $V_2 = V_3 = 0.05$, except for HBT radii for which all $V_n$ vanish.
In contrast, we let the anisotropy parameter $\xi$ vary, giving it values from $-0.5$ to 0 in steps of $0.1$, together with  0.15 and 0.3.
According to our argumentation in the previous section, these positive values, which lead to smaller pressure in the radial direction as perpendicular to it, should not be relevant for freeze out; 
yet we included them for reference sake. 

\begin{figure}[b!]
	\includegraphics*[width=\linewidth]{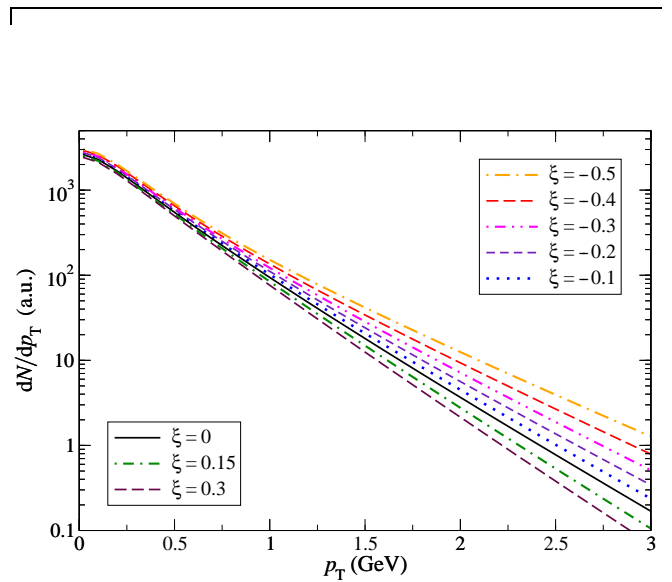}
	\caption{\label{fig:dNdpT_vs_xi} Transverse spectra for fixed $\Lambda$ and varying anisotropy parameter $\xi$.}
\end{figure}
\begin{figure*}[t!]
	\includegraphics*[width=0.495\linewidth]{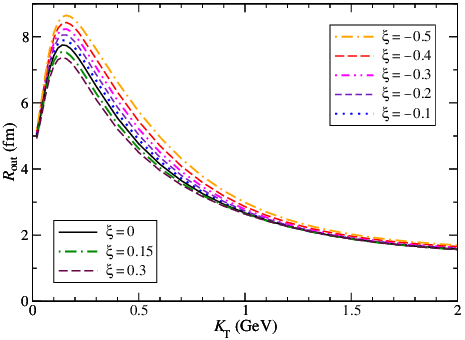}\hfill
	\includegraphics*[width=0.495\linewidth]{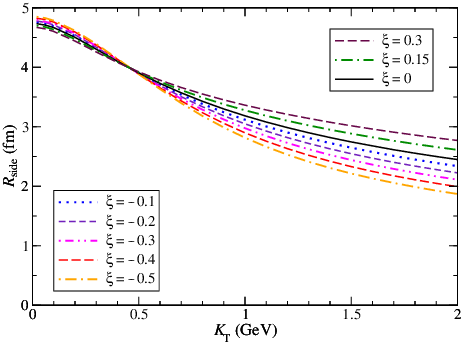}%
	
	\includegraphics*[width=0.495\linewidth]{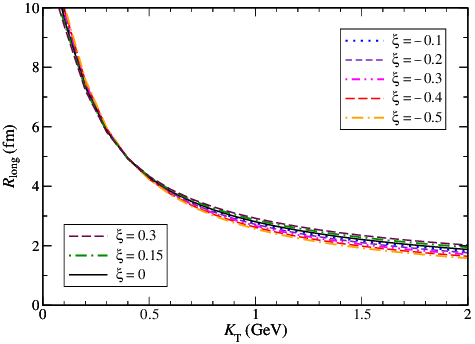}\hfill
	\includegraphics*[width=0.495\linewidth]{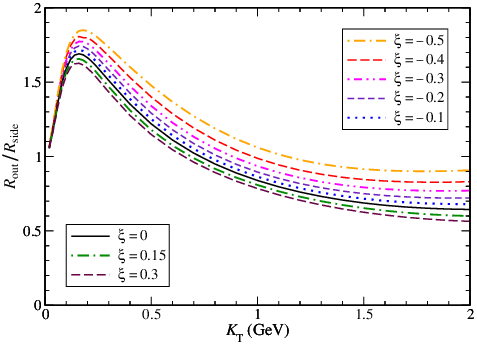}%
	
	\vspace{-2mm}
	\caption{\label{fig:HBT-R_vs_xi} HBT radii for fixed $\Lambda$ and varying anisotropy parameter $\xi$. Top left: $R_{\rm out}$, top right: $R_{\rm side}$, bottom left: $R_{\rm long}$, bottom right: ratio $R_{\rm out}/R_{\rm side}$.}\vspace{-3mm}
\end{figure*}
Figure~\ref{fig:dNdpT_vs_xi} shows the resulting transverse momentum distributions.
As is to be expected, non-zero values of $\xi$ lead to deviations from the almost exponential shape valid in the isotropic case. 
More precisely, the spectrum becomes harder when $\xi$ goes to increasingly negative values. 
This clearly reflects the growing radial pressure---or equivalently effective radial temperature $\Lambda/\sqrt{1+\xi}$---obtained by assuming $\xi<0$. 
In figure~\ref{fig:HBT-R_vs_xi}, we display the various HBT radii, together with the ratio $R_{\rm out}/R_{\rm side}$, as functions of the pair transverse momentum $K_T$. 
To be more precise, the radii $R_{\rm side}^2$ and $R_{\rm long}^2$ are the $f_{\rm an.}$-weighted averages over the freeze-out hypersurface of $y^2=r^2\sin^2\phi$ and $z^2=\tau^2\sinh^2\varsigma$, respectively, while $R_{\rm out}^2$ is the average of $(x-K_T t/E_{\bm{K}})^2$, where $x=r\cos\phi$ and $t=\tau\cosh\varsigma$.

As was just mentioned, negative values of $\xi$ amount to a larger ``radial temperature'', and thus to higher thermal velocities in the outwards direction.
Since at the same time the emission duration barely changes, this naturally leads to a larger $R_{\rm out}$, as observed in the upper left panel, as well as to a larger ratio $R_{\rm out}/R_{\rm side}$ (lower right panel) 
In turn, the longitudinal radius $R_{\rm long}$ shown in the lower left panel is to a large extent unaffected by $\xi$; this could be anticipated since the longitudinal part of the occupation factor remains unchanged. 
On the other hand, the behaviour of the sidewards radius $R_{\rm side}$ with varying $\xi$ seen in the upper right panel of figure~\ref{fig:HBT-R_vs_xi} is more involved, and we did not find a satisfactory explanation describing all its details. 

\begin{figure}[b!]
	\includegraphics*[width=\linewidth]{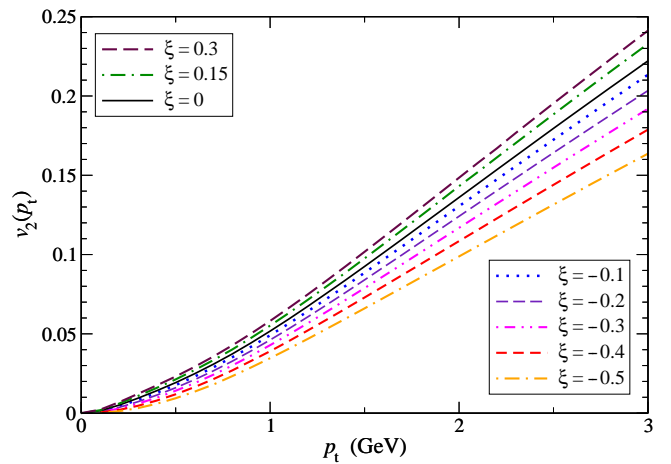}\vspace{-2mm}
	\caption{\label{fig:v2_vs_xi} Elliptic flow $v_2(p_t)$ for fixed $\Lambda$ and varying anisotropy parameter $\xi$.}\vspace{-1.25mm}
\end{figure}
The transverse-momentum dependence of elliptic flow $v_2$ for various $\xi$ values is shown in figure~\ref{fig:v2_vs_xi}; 
triangular flow $v_3$ follows exactly the same trend, so that we do not show it.  
Thus, anisotropic flow decreases when $\xi$ becomes more negative, that is, as the radial temperature grows. 
This behaviour reflects the fact that an increase in random thermal motion tends to dilute the effect of directed collective behaviour encoded in the flow velocity and its anisotropies, i.e., it diminishes the $v_n$ values, as seen here.  
\begin{figure}[t!]
	\includegraphics*[width=\linewidth]{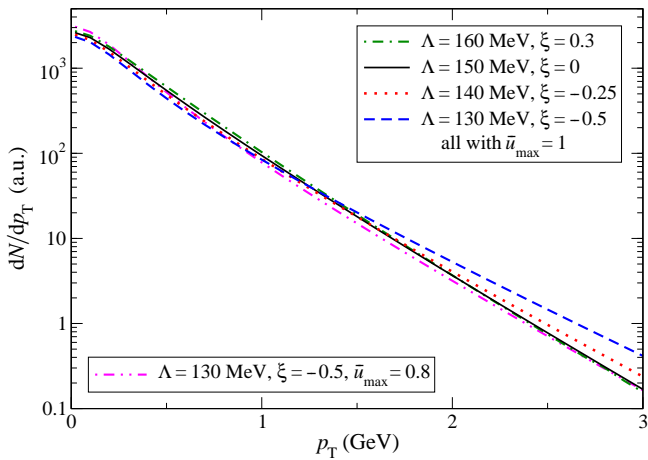}\vspace{-2mm}
	\caption{\label{fig:dNdpT_for_various_T-and-xi} Transverse spectra for various choices of $\Lambda$ and $\xi$.}\vspace{-3mm}
\end{figure}

\begin{figure}
	\includegraphics*[width=\linewidth]{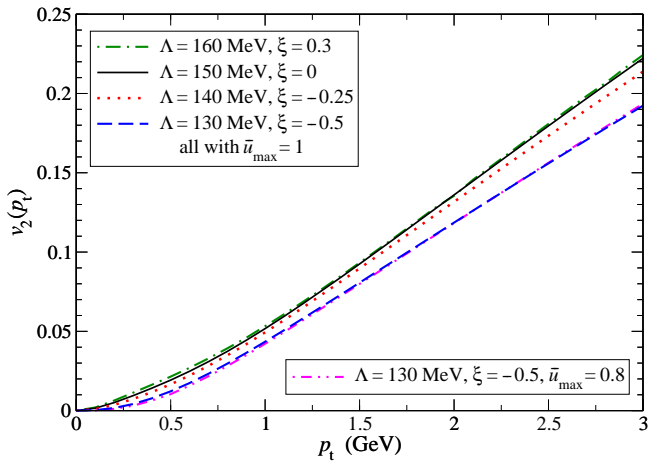}\vspace{-2mm}
	\caption{\label{fig:v2_for_various_T-and-xi} Elliptic flow for various choices of $\Lambda$ and $\xi$.}\vspace{-2mm}
\end{figure}

Before going any further, let us note that in a more complete approach, the local anisotropy parametrized in this work by $\xi$ should not be uniform, but rather position-dependent. 
In particular, $\xi$ (or similar parameters) would normally be function of the azimuthal angle $\phi$, paralleling the corresponding dependence of the velocity profile, as we now argue.\footnote{Similarly, $\Lambda$ also might depend on $\phi$, yet we leave this possibility aside to simplify the discussion.} 
The fluid-particle conversion, whose modelling $\xi$ is supposed to facilitate, roughly happens when the fluid expansion rate $\nabla_\mu u^\mu(\textsf{x})$ becomes comparable to that of elastic scatterings. 
Since the flow velocity varies with $\phi$, so does the expansion rate, which motivates an azimuthal dependence of $\xi$. 
On the other hand, the scattering rate depends on particle density, obtained by integrating the occupancy factor over momentum, and on the relative velocity of particles.  
As follows from a straightforward change of integration variable~\cite{Rybczynski:2012ee}, the density is inversely proportional to $\sqrt{1+\xi(\textsf{x})}$, thus a priori $\phi$-dependent. 
In turn, the typical relative velocity is controlled by the (effective) temperature(s) of the decoupling medium, thus function of $\phi$ as well\ldots
All in all, every relevant physical quantity depends on azimuth, so it is non-trivial---and within the scope of this paper rather academic---to determine the actual dependence of $\xi$. 
In any case, there will be such a dependence, which will affect the anisotropic flow coefficients $v_n$. 
The results shown in figures~\ref{fig:v2_vs_xi} and~\ref{fig:v2_for_various_T-and-xi} are thus to be taken with a grain of salt, since they neglect this ingredient.

After having investigated the influence of $\xi$ when all other parameters are fixed, we now want to illustrate the degeneracy introduced by this new parameter, showing that very similar values of the observables can be obtained with different pairs $(\Lambda,\xi)$. 
Note that we did not attempt to optimize the results we now report by fine tuning the parameters, as will be made apparent by the values of the latter. 

\begin{figure}[b!]
	\includegraphics*[width=\linewidth]{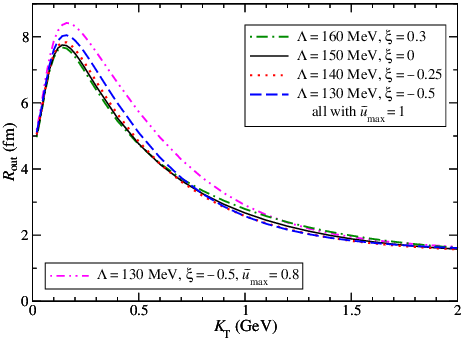}
	
	\includegraphics*[width=\linewidth]{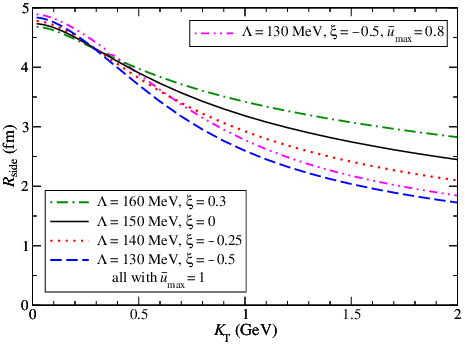}
	\vspace{-5mm}
	
	\caption{\label{fig:HBT-R_for_various_T-and-xi} HBT radii $R_{\rm out}$ (top panel) and $R_{\rm side}$ (bottom panel) for various choices ($\Lambda$,$\xi$).}
\end{figure}
In figure~\ref{fig:dNdpT_for_various_T-and-xi}, we display the transverse momentum spectra for four sets of values of $(\Lambda,\xi)$, with $\Lambda$ varying between 130 and 160\;MeV and $\xi$ ranging from $-0.5$ to 0.3. 
In all four cases, the values of all other parameters are the same as above, in particular $\bar{u}_{\max}=1$.
All four curves are barely distinguishable below $p_T=1.5\,$GeV, above which that with ($\Lambda=130\,$MeV, $\xi=-0.5$) starts curving up. 
The spectrum for ($\Lambda=140\,$MeV, $\xi=-0.25$) only starts differing from those with larger $\Lambda$ from about 2\;GeV onwards, while the remaining two stay very close up to at least 3\,GeV. 
In addition, we show in the same figure the spectrum for ($\Lambda=130\,$MeV, $\xi=-0.5$) and a different flow velocity, namely with $\bar{u}_{\max}=0.8$. 
The change in $\bar{u}_{\max}$ makes the spectrum almost collapse on that for ($\Lambda=150\,$MeV, $\xi=0$), with at most a 15\% relative difference over the whole momentum range. 

The HBT radii $R_{\rm out}$ and $R_{\rm side}$ and the elliptic flow $v_2$ for the same sets of parameters as in figure~\ref{fig:dNdpT_for_various_T-and-xi} are respectively shown in figures~\ref{fig:HBT-R_for_various_T-and-xi} and \ref{fig:v2_for_various_T-and-xi}. 
As in the case of the transverse spectra, the values of $R_{\rm out}$ or $v_2$ for all four pairs $(\Lambda,\xi)$ in the case $\bar{u}_{\max}=1$ are very close to each other, with ($\Lambda=130\,$MeV, $\xi=-0.5$) being most apart from the other three. 
We also include the result of the computation with $\bar{u}_{\max}=0.8$ which gives a good approximation to the $p_T$-distribution: for $v_2$, it basically makes no difference with respect to the case $\bar{u}_{\max}=1$, whereas the departure is more marked for $R_{\rm out}$. 

All in all, the results for transverse-momentum distributions, $R_{\rm out}$, and $v_2$ support our claim that introducing an extra parameter opens a much wider range for the ``freeze-out temperature'', here $\Lambda$, without affecting drastically the values of the observables. 

In contrast, the sidewards HBT radius $R_{\rm side}$ displayed in the bottom panel of figure~\ref{fig:HBT-R_for_various_T-and-xi} is much more sensitive to the choice of decoupling parameters $(\Lambda,\xi)$. 
This is actually somewhat reassuring, since femtoscopic measurements are precisely designed to probe the space-time configuration at decoupling~\cite{Lisa:2008gf}.

\section{Discussion}
\label{s:discussion}

We have argued that there are two main motivations for resorting to an anisotropic momentum distribution to describe the transition from usual dissipative fluid dynamics to a particle description at the end of the evolution of the fireball created in ultrarelativistic heavy-ion collisions. 
Firstly, this ansatz is supported by non-relativistic studies of freeze out~\cite{Hamel:1966}.
Secondly, this could help diminish the sensitivity of computed observables on the parameters introduced by the decoupling prescription, and thus lead to a smoother matching between models, in the spirit of seeing fluid dynamics emerging as the effective theory of some underlying, more microscopic dynamics. 

As a matter of fact, our findings for transverse spectra, $R_{\rm out}$, and $v_2$ (figures~\ref{fig:dNdpT_for_various_T-and-xi}--\ref{fig:v2_for_various_T-and-xi}) support the idea that introducing an extra parameter, which governs the local momentum anisotropy at decoupling, opens a much wider range for the ``freeze-out temperature'', here $\Lambda$, without changing significantly the values of the observables. 
This is admittedly not too surprising, since we introduced one new degree of freedom. 
Yet at the risk of repeating ourselves, it emphasizes the fact that the ``freeze-out temperature'' is just a parameter for switching between two models, not a real physical temperature determined by some ``critical''---in a loose sense---energy or entropy density for which the medium properties change drastically. 
Being such a parameter---like say a renormalization scale---, it may not have a dramatic impact on measurable quantities. 

Accordingly, it seems possible to find a whole region of parameters to which the ``early time'' signals like anisotropic flow---which carry information on the properties of the fireball along its whole evolution~\cite{Heinz:2013th}, rather than on decoupling itself---are to a large extent insensitive. 
On the other hand, some sensitivity remains for the observables which are governed by the freeze-out process. 

In the present exploratory study, we postulated the asymmetric form of the occupation factor at decoupling $f_{\textrm{an.}}$, and investigated some of the consequences within a toy model. 
The actual form of $f_{\textrm{an.}}$, together with that of the associated hydrodynamical quantities, still has to be calculated in a more microscopic approach~\cite{BFL_inprep}. 
This involves at the same time a discussion of the freeze-out hypersurface $\Sigma$, whose position in space-time obviously depends on the amount of momentum anisotropy in the phase-space distribution. 

Once this is done, it will be necessary to study how the ``improved'' prescription can be implemented in practice, i.e. how numerical simulations of dissipative fluid dynamics, anisotropic hydrodynamics, and particle transport can be glued together in a satisfactory manner. 
An important point will be to check what the shortcomings of the sudden freeze-out scenario, in particular the backflow of particles through $\Sigma$~\cite{Pratt:2014vja}, become in the new approach: 
if there is more freedom in choosing the decoupling hypersurface, some choices may be more convenient than others. 
Eventually, it will be interesting to investigate the possible relation of the new prescription, which in essence still assumes a sudden fluid-particle conversion, with continuous emission~\cite{Grassi:1994nf&ng,Sinyukov:2002if}. 
For instance, one may wonder if it is possible to mimic the latter within the former, or whether one has to formulate a continuous version of the ``anisotropic decoupling'' scenario.


\end{document}